\documentclass[a4paper]{article}

\usepackage[utf8]{inputenc}
\usepackage[francais,american]{babel}
\usepackage{enumerate}
\usepackage{amsfonts,amsmath,amssymb,amsthm,amstext,latexsym}	
\usepackage{hyperref}
\usepackage{xspace}
\usepackage{subcaption}
\usepackage{array}
\usepackage{todonotes}
\usepackage{ifthen}
\usepackage{intcalc}
\usepackage{stmaryrd}
\usepackage{fullpage}

\newcommand{\ema}[1]{\ensuremath{#1}\xspace}
\newcommand{\mtbf}{\ema{\mu}}
\newcommand{\nodes}{\ema{N}}
\newcommand{\mtbfind}{\ema{\mu_{\text{ind}}}}

\newcommand{\ccc}{\ema{C}}
\newcommand{\rrr}{\ema{R}}
\newcommand{\ddd}{\ema{D}}

\newcommand{\period}{T}
\renewcommand{\time}[1][]{\ema{\mathcal{T}_{\text{#1}}}}

\newcommand{\tbase}{\time[base]}
\newcommand{\tff}{\time[ff]}
\newcommand{\tfails}{\time[fails]}
\newcommand{\tfin}{\time[final]}
\newcommand{\tCal}{\time[Cal]}
\newcommand{\tIO}{\time[I/O]}
\newcommand{\tDown}{\time[Down]}
\newcommand{\twork}{\time[re-exec]}
\newcommand{\toptT}{\ema{\time[Time]^{\text{opt}}}}
\newcommand{\toptE}{\ema{\time[Energy]^{\text{opt}}}}

\newcommand{\power}[1][]{\ema{\mathcal{P}_{\text{#1}}}}
\newcommand{\pCal}{\power[Cal]}
\newcommand{\pIO}{\power[I/O]}
\newcommand{\pDown}{\power[Down]}
\newcommand{\pIdle}{\power[Static]}
\newcommand{\pStatic}{\power[Static]}

\newcommand{\E}[1][]{\ema{\mathcal{E}_{\text{#1}}}}
\newcommand{\efin}{\E[final]}

\newcommand{\workduringckpt}{\omega}
\newcommand{\cosi}{\alpha}
\newcommand{\ios}{\beta}
\newcommand{\dos}{\gamma}
\newcommand{\ratio}{\rho}

\newcommand{\algoE}{\textsc{AlgoE}\xspace}
\newcommand{\algoT}{\textsc{AlgoT}\xspace}

 \author{Guillaume Aupy$^{1}$, Anne Benoit$^{1}$, Thomas H\'erault$^{2}$,\\
 Yves Robert$^{1,2}$, and Jack Dongarra$^{2}$\\
 $1.$ LIP,  \'Ecole Normale Sup\'erieure de Lyon, CNRS \& INRIA, France\\
 $2.$ University of Tennessee Knoxville, USA
 }

\title{Optimal Checkpointing Period: Time vs. Energy}

\begin{document}
\maketitle

\begin{abstract}
This short paper deals with parallel scientific applications using non-blocking and periodic coordinated checkpointing
to enforce resilience. We provide a model and detailed formulas for total execution time
and consumed energy. We characterize the optimal period for both objectives, and we assess
the range of time/energy trade-offs to be made by instantiating the model 
with a set of realistic scenarios for Exascale systems. We give a particular emphasis to I/O transfers,
because the relative cost of communication is
expected to dramatically increase, both in terms of latency and consumed energy, for future Exascale platforms.
\end{abstract}

\section{Introduction}
\label{sec.intro}

A significant research effort is focusing on the
characteristics, features, and challenges of High Performance
Computing (HPC) systems capable of reaching the Exaflop performance
mark~\cite{IESP-Exascale,DARPA}. The portrayed Exascale systems will
necessitate billion way parallelism, resulting not only in a massive
increase in the number of processing units (cores), but also in terms
of computing nodes. Considering the relative slopes describing the evolution of the
reliability of individual components on one side, and the evolution of
the number of components on the other side, the reliability of the
entire platform is expected to decrease, due to probabilistic
amplification.  Even if each independent component is quite reliable, the Mean Time Between Failures (MTBF) is expected to drop drastically. Executions of large parallel applications on these
systems will have to tolerate a higher degree of errors and failures
than in current systems. The de-facto general-purpose error recovery technique in high performance computing 
is checkpoint and rollback recovery. Such protocols employ checkpoints to periodically save 
the state of a parallel application, so that when an error strikes some process, 
the application can be restored into one of its former states. The most widely used protocol is coordinated 
checkpointing, where all processes periodically stop computing and synchronize to write critical application
data onto stable
storage. Coordinated 
checkpointing is well understood, at least in its blocking form (when no computing activity takes place
during checkpoints), and good
approximations of the optimal checkpoint interval exist; they are known as
Young's and Daly's formula~\cite{young74,daly04}. 

While reliability is a major concern for Exascale, another key challenge is to minimize 
energy consumption, both for economic and environmental reasons. 
One of the most power-consuming components of today's systems is the processor: 
even when idle, it dissipates a significant fraction of the total power. However, for future
Exascale systems, the power dissipated to execute I/O transfers
is likely to play an even more important role, because the relative cost of communication is
expected to dramatically increase, both in terms of latency and consumed energy~\cite{Shalf2011}. 

In this short paper, we investigate trade-offs between execution time and energy consumption
for the execution of parallel applications on future Exascale systems. 
The optimal period $\toptT$ given by Young's and Daly's formula~\cite{young74,daly04} 
will minimize (expected) execution time. But will it minimize energy consumption?
The answer is negative, mainly because the fraction of power \pCal spent when computing (by the CPUs) is not 
the same as the fraction of power \pIO spent when checkpointing. 
In particular, we revisit the work of Meneses, Sarood and Kalé~\cite{Kale2012} for checkpoint/restart, where formulas 
are given to compute the time-optimum and energy-optimum periods. 
However, our model is more precise: (i) we carefully assess  the impact of the
power consumption required for I/O activity, which is likely to play a key role
at the Exascale; (ii) we consider non-blocking checkpointing that can be partially
overlapped with computations; (iii) we give a more accurate analysis of the consumed energy.

Altogether, this short paper provides the following main contributions:
\begin{itemize}
\item We provide a refined analytical model to compute both the execution time and the
consumed energy with a given checkpoint 
period. The model handles the case where checkpointing activity can be non-blocking, i.e.,  
partially overlapped with computations. 
\item We provide analytical formulas to approximate the optimal period for time $\toptT$
 as well as the optimal period  for energy $\toptE$,  thereby refining and extending
Daly~\cite{daly04} and Meneses, Sarood and Kalé~\cite{Kale2012} results to
non-blocking checkpoints.
\item We assess the range of time/energy trade-offs to be made by instantiating the model 
with a set of realistic scenarios for Exascale systems. 
\end{itemize}

\section{Model}
\label{sec.model}

In this section, we introduce all the model parameters. We
start with parameters related to resilience (checkpointing)
before moving to parameters related to
energy consumption.

\subsection{Checkpointing}

We model coordinated checkpointing~\cite{CL85} where 
checkpoints are taken at regular intervals, after some fixed amount of
work units have been performed. This corresponds to an 
execution partitioned into periods of duration $\period$. Every period, 
a checkpoint of length $\ccc$ is taken.

An important question is
whether checkpoints are blocking or not.  On some architectures, we
may have to stop executing the application before writing to the
stable storage where the checkpoint data is saved; in that case
checkpoint is fully blocking.  On other architectures, checkpoint data
can be saved on the fly into a local memory before the checkpoint is
sent to the stable storage, while computation can resume progress; in
that case, checkpoints can be fully overlapped with computations.  To
deal with all situations, we introduce a slow-down factor
$\workduringckpt$: during a checkpoint of duration $\ccc$, the work
that is performed is $\workduringckpt \ccc$ work units. In other words,
$(1-\workduringckpt) \ccc$ work units are wasted due to checkpoint
jitter disrupting the progress of computation.  Here, $0 \leq
\workduringckpt \leq 1$ is an arbitrary parameter.  The case
$\workduringckpt =0$ corresponds to a fully blocking checkpoint, while
$\workduringckpt=1$ corresponds to a checkpoint totally overlapped with computations.
All intermediate situations can be represented.

Next we have to account for failures. During $t$ time units of
execution, the expectation of the number of failures is
$\frac{t}{\mtbf}$, where \mtbf is the MTBF (Mean Time Between Failures) of
the platform.  Note that if the platform if made of $\nodes$ identical
resources whose individual mean time between failures is \mtbfind,
then $\mtbf = \frac{\mtbfind}{\nodes}$. This relation is agnostic of
the granularity of the resources, which can be anything from a single
CPU to a complex multi-core socket. When a failure strikes,
there is a downtime of length \ddd (time to reboot the resource or set
up a spare), and then a recovery of length \rrr (time to read the last stored checkpoint).
The work executed by the application since the last checkpoint and before the failure needs 
to be re-executed. Clearly, the shorter the period $\period$, the less work to re-execute, but also
the more overhead due to frequent checkpoints in a failure-free execution. The best trade-off 
when $\workduringckpt = 0$ (blocking checkpoint) is achieved for $\period = \sqrt{2 \ccc \mu} + \ccc$
(Young's formula~~\cite{young74}) or $\period = \sqrt{2 \ccc (\mu +\ddd+ \rrr)} + \ccc$ 
(Daly's formula~\cite{daly04}).
Both formulas are first-order approximations and valid only if all checkpoint parameters \ccc, \ddd and \rrr
are small in front of $\mu$ (and these formulas collapse if they become negligible).
In Section~\ref{sec.period}, we show how to extend these formulas to the case of non-blocking checkpoints (see
also~\cite{ccpe-2012-ckpt} for more details).

\subsection{Energy}
\label{sec.model.energy}

To compute the energy consumption of the application, we need to consider the energy 
consumption of the different phases, and hence the power consumption at each time-step. 
To this purpose,  we define:
\begin{itemize}
	\item {\bf \pIdle:} this is the base power consumed when the platform is switched on.
	\item {\bf \pCal:} when the platform is active, we have to consider the CPU overhead in addition to the static power \pIdle.  
	\item {\bf \pIO:} similarly, this is the power overhead due to file I/O. This supplementary power   
consumption is induced by checkpointing, or when recovering from a failure.
	\item {\bf \pDown:} for coordinated checkpointing, when one processor fails, the rest of the
machine stays idle. \pDown is the power consumption overhead when one machine is down, that may be incurred for instance by rebooting the machine. In general, we let $\pDown=0$.
\end{itemize}

Meneses, Sarood and Kalé~\cite{Kale2012} have a simpler model with two parameters, namely $L$, the base power
(corresponding to \pIdle with our notations), and $H$, the maximum power 
(corresponding to $\pIdle+\pCal$ with our notations). They use $\pIO = \pDown=0$.

In Section~\ref{sec.period}, we show how to compute the optimal period that minimizes  the energy 
consumption.  In Section~\ref{sec.experiments},  we instantiate the model with expected values for power 
consumption of Exascale platforms.

\section{Optimal checkpointing period}
\label{sec.period}

We consider a parallel application whose execution time is 
\tbase without any overhead due to the resilience method or the occurrence of failures.
We compute the expectation \tfin of the total execution time (accounting both for checkpointing and for failures)
in Section~\ref{sec.tfin}, and the expectation \efin of the total energy consumed during this execution
of length \tfin in Section~\ref{sec.efin}. We will compute the optimal period $T$ that minimizes
the objective, either \tfin or \efin.

\subsection{Execution time}
\label{sec.tfin}

The total execution time \tfin of the application depends on two sources of overhead. We 
first compute \tff, the time taken by a fault-free execution, thereby accounting only 
for the overhead due to 
periodic checkpointing. Then we compute \tfails, the time lost due to failures. Finally,  
$\tfin=\tff+\tfails$. We detail here both computations:
\begin{itemize}
	\item The reasoning to derive \tff is simple. We need to execute a total amount of work equal to~\tbase.
	During each period of length~$\period$,
there is an amount of time $\period - \ccc$ where only computations take place, and an amount of time~$\ccc$ 
of checkpointing, where only a work $\workduringckpt \ccc$ is done. Therefore, the total number of work
units executed during a period of length~$\period$ is $\period - \ccc +\workduringckpt \ccc = \period - (1- \workduringckpt) \ccc$,
and \\[-.2cm]
\[\tff = \tbase \frac{\period}{\period - (1- \workduringckpt) \ccc}.\]
	\item The reasoning to compute \tfails is the following. Since the mean time between two failures is $\mu$, 
	the average number of failures during execution is $\frac{\tfin}{\mu}$.
For each failure, the time lost is expressed as: 
	\begin{itemize}
		\item $\ddd + \rrr$ for downtime and recovery; 
	  	\item a time $\workduringckpt \ccc$ for the work that was done during the previous 
		checkpoint and that has to be redone because it was not checkpointed (because of the failure); 
		\item with probability $\frac{\period - \ccc}{\period}$, the failure happens while we 
	are not checkpointing, and the time lost is on average~$A = \frac{\period - \ccc}{2}$; 
		\item otherwise, with probability $\frac{\ccc}{\period}$, the failure happens while we are
	checkpointing, and the time lost is on average
	$B = \period-\ccc + \frac{\ccc}{2} = \period-\frac{\ccc}{2}$.
	\end{itemize}
	The time lost for each failure is 
	$$\ddd + \rrr +\workduringckpt\ccc + \frac{\period - \ccc}{\period} A + \frac{\ccc}{\period} B = \ddd + \rrr +\workduringckpt\ccc + \frac{\period}{2}.$$
	Finally, \\[-.7cm]
	\begin{align*}
	\tfails &= \frac{\tfin}{\mu} \left ( \ddd + \rrr +\workduringckpt\ccc + \frac{\period}{2} \right ) .
	\end{align*}
\end{itemize}

We are now ready to express the total execution time: \\[-.5cm]
\begin{align*}
\tfin 
&= \tff + \tfails \\
 &= \tbase\frac{\period}{\period - (1- \workduringckpt)\ccc} + \frac{\tfin}{\mu}\left ( \ddd + \rrr + \workduringckpt \ccc + \frac{\period}{2} \right ) \\
 &= \frac{\period}{\left ( \period- (1-\workduringckpt)\ccc \right ) \left ( 1- \frac{\ddd + \rrr + \workduringckpt \ccc + \period /2}{\mu}\right )}\tbase\\
 &= \frac{\period}{\left ( \period- a \right ) \left ( b - \frac{\period}{2\mu} \right )}\tbase ,
\end{align*}
where $a= (1-\workduringckpt)\ccc$ and $b=  1- \frac{\ddd + \rrr + \workduringckpt \ccc }{\mu}$.  

\smallskip
This equation is minimized for  
\begin{equation}
\toptT = \sqrt{2(1-\workduringckpt)\ccc(\mu - (\ddd + \rrr +\workduringckpt \ccc))}. 
\label{sol.for.time}
\end{equation}

When $\workduringckpt=0$, we obtain an expression close to that of Young and Daly,
but slightly different because they have less accurately approximated the total execution time.  
In the following, we let \algoT be the checkpointing strategy that checkpoints with period \toptT.

\subsection{Energy consumption}
\label{sec.efin}

In order to compute the total energy consumption of the execution, we consider the different phases
during which the different powers introduced in Section~\ref{sec.model.energy} are used: 
\begin{itemize}
	\item First,  we consume \pStatic during each time-step of the execution.
Indeed, even when a node fails and is shutdown, we still pay for the power of all the 
other nodes, for the cooling system, etc. The corresponding energy cost is $\tfin \pIdle$.
\smallskip
	\item Next, let \tCal be the time during which the CPU is used, inducing a power overhead \pCal. 
	\tCal 	includes the base work \tbase, and 
 \twork, the work that must be re-executed after each failure (which we multiply 
		by the number of failures $\tfin / \mu$):
		
		 \begin{itemize}
			\item with probability $\frac{\period - \ccc}{\period}$, the failure does not happen
		during a checkpoint, and  the work to re-execute is
		 $A = \workduringckpt \ccc + \frac{\period - \ccc}{2}$; 
			\item with probability $\frac{\ccc}{\period}$, the failure happens during the
		execution of a checkpoint, and the work to re-execute is
		 $B = \workduringckpt \ccc + \period - \ccc + \frac{\workduringckpt \ccc}{2}$.
		\end{itemize}
		\smallskip
		We derive $\twork = \frac{\period - \ccc}{\period} A + \frac{\ccc}{\period} B$, hence\\[-.2cm]
		$$\twork =  \workduringckpt \ccc + \frac{\period^2 - \ccc^2}{2\period} + \frac{\workduringckpt \ccc^2}{2\period}.$$
Finally, we have: 
\[ \tCal = \tbase + \frac{\tfin}{\mu}\left ( \workduringckpt \ccc + \frac{\period^2 - \ccc^2}{2\period} + \frac{\workduringckpt \ccc^2}{2\period}\right ) . \]
The corresponding energy consumption is $\tCal \pCal$.
\smallskip
	\item Let \tIO be the time during which the I/O system is used,  inducing a power overhead \pIO. This time corresponds to
checkpointing and recovery from failures. 
	\begin{itemize}
		\item The total number of checkpoints that are taken in a fault-free execution  is equal 
		to the number of periods, 
	$\frac{\tbase}{\period - (1-\workduringckpt)\ccc}$, and the time taken by checkpoints
	is therefore $\frac{\tbase C}{\period - (1-\workduringckpt)\ccc}$. 
	\smallskip
		\item For each failure, there is an additional overhead:
		\begin{enumerate}
			\item the system needs to recover, which lasts \rrr time-steps;
			\item with probability $\frac{\period - \ccc}{\period}$, the failure does not happen
		during a checkpoint, and there is no additional I/O overhead;
			\item however, with probability $\frac{\ccc}{\period}$, the failure happens during a 
		checkpoint, and the I/O time wasted is (in average)  $\frac{\ccc}{2}$.
		\end{enumerate}
	\end{itemize}
Altogether, we obtain
\[\tIO = \frac{\tbase \ccc}{\period - (1-\workduringckpt)\ccc} + \frac{\tfin}{\mu}\left ( \rrr + \frac{\ccc^2}{2\period}\right ) .  \]
The corresponding energy consumption is $\tIO \pIO$.
\smallskip
	\item  Finally, let \tDown be the total down time, incurring a power overhead \pDown.
	We have 
\[\tDown = \frac{\tfin}{\mu}\ddd ,\]
and the corresponding energy cost is $\tDown \pDown$. This term is only included for full generality, as 
we expect to have $\pDown=0$ in most scenarios.
\end{itemize}

\medskip
The final expression for the total energy consumed is
\begin{align*}
\efin &= \tCal \pCal + \tIO \pIO + \tDown \pDown + \tfin \pIdle\\
 &= \left ( \tbase + \frac{\tfin}{\mu}\left ( \workduringckpt \ccc + \frac{\period^2 - \ccc^2}{2\period} + \frac{\workduringckpt \ccc^2}{2\period} \right ) \right ) \pCal \\
 & \quad+ \left ( \frac{\tfin}{\mu}\left ( \rrr + \frac{\ccc^2}{2\period}\right ) + \ccc \frac{\tbase}{\period - (1-\workduringckpt)\ccc}\right ) \pIO \\
 & \quad + \frac{\tfin}{\mu}\ddd \pDown + \tfin \pStatic .
\end{align*}

It is important to understand that $\tfin \neq \tCal + \tIO + \tDown$, unless $\workduringckpt=0$.
Indeed, CPU and I/O activities are overlapped (and both consumed) when checkpointing.
To ease the derivation  of the optimal period that minimizes \efin, we introduce some notations
and let  $\pCal =\cosi \pStatic$, $\pIO = \ios \pStatic$, 
and $\pDown = \dos \pStatic$.
Re-using parameters $a= (1-\workduringckpt)\ccc$ and $b=  1- \frac{\ddd + \rrr + \workduringckpt \ccc }{\mu}$
from Section~\ref{sec.tfin}, we obtain:

\[
\frac{\tfin'}{ \tbase} = \frac{-ab + \frac{T^2}{2\mu}}{\left ( \period - a \right )^2\left (b - \frac{\period}{2\mu} \right)^2} ,
\text{~~~~~and}\]
\begin{align*}
 \frac{\efin'}{\pStatic} &\!\!\!=\! \frac{\tfin'}{\mu}\! \left ( \cosi \workduringckpt \ccc \!+\! \ios \rrr\!+\! \dos \ddd \!+\! \frac{\cosi \period}{2} \!-\! \frac{\cosi(1-\workduringckpt) \ccc^2}{2\period}\!+\! \frac{\ios \ccc^2}{2\period} \!+\! \mu \right ) \\
&+ \frac{\tfin}{2\mu} \left ( \cosi \!+\! \frac{\cosi(1-\workduringckpt)\ccc^2}{\period^2} \!-\! \frac{\ios \ccc^2}{\period^2} \right ) \! -\! \frac{\ios \ccc \tbase}{\left ( \period \!-\! (1\!-\!\workduringckpt)\ccc \right )^2} .
\end{align*}

Then, letting $K = \frac{\left ( \period - a \right )^2\left (b - \frac{\period}{2\mu} \right)^2}{\pStatic\tbase}$, we have:

%
\begin{align*}
K\efin' 
&= \frac{-ab \!+\! \frac{\period^2}{2\mu}}{\mu} \!\!\left ( \!\left (\cosi \workduringckpt \ccc 
\!+\! \ios \rrr\!+\! \dos \ddd \!+\! \mu\right ) \! +\! \frac{\cosi \period}{2} \!+\! \frac{\cosi(1-\workduringckpt) \ccc^2}{2\period}
\!+\! \frac{\ios \ccc^2}{2\period}\! \right ) \\
& \quad+ \frac{\left ( \period \!-\! a \right )\!(b \!-\! \frac{\period}{2\mu})}{2\mu}\!\! \left ( \cosi \!+\! \frac{\cosi(1-\workduringckpt)\ccc^2 \!-\! \ios\ccc^2}{\period} \right ) \!-\! \ios \ccc \left (b \!-\! \frac{\period}{2\mu} \right)^2 \\
&= \period^3 \!\!\left (\!\frac{1}{4\mu} \!-\!\frac{1}{4\mu}\! \right ) \!+\! \period^2\!\! \left ( \frac{\cosi \workduringckpt \ccc \!+\! \ios \rrr \!+\! \dos \ddd}{2\mu^2} \!+\! \frac{b \!+\!\frac{a}{2\mu}}{2\mu} \!-\! \frac{\ios \ccc}{4\mu^2} \!+\!\frac{1}{2\mu}\right )  \\
& \quad+ \period \left ( \!-\!\frac{ab}{2\mu}\! -\!\frac{ab}{2\mu} \!+\!\frac{\ios \ccc b}{\mu}  \!-\! 2\frac{( \cosi(1-\workduringckpt) \!-\! \ios) \ccc^2}{4\mu^2} \right )  \!-\! \ios \ccc b^2\\
&\quad - \frac{ab \left (\cosi \workduringckpt \ccc \!+\! \ios \rrr \!+\! \dos \ddd \!+\! \mu \right )}{\mu} \!-\!\left ( \frac{b}{2\mu}\!-\!\frac{a}{4\mu^2}\right )\!(\cosi(1-\workduringckpt)\!-\!\ios)\ccc^2 \\
& \quad + \frac{1}{\period}\!\! \left (\left ( \cosi(1-\workduringckpt)\!-\! \ios\right ) \frac{\ccc}{2\mu} \!-\! \left ( \cosi(1-\workduringckpt)\!-\! \ios\right ) \frac{\ccc}{2\mu} \right )\\
&= \period^2 \left (\frac{\cosi \workduringckpt \ccc \!+\! \ios \rrr \!+\! \dos \ddd}{2\mu^2} \!+\! \frac{b}{2 \mu} \!+\! \frac{a \!-\! \ios \ccc}{4\mu^2} \!+\!\frac{1}{2\mu}\right ) \\
&\quad + \period \left ( \frac{(\ios \ccc\!-\!a) b}{\mu}\!-\! 2\frac{(\cosi(1-\workduringckpt) \!-\! \ios) \ccc^2}{4\mu^2}  \right ) \\
&\quad  - \frac{ab \left ( \cosi\workduringckpt \ccc \!+\! \ios \rrr \!+\! \dos \ddd \!+\! \mu \right  )}{\mu}\! - \!\ios \ccc b^2 \\
&\quad + \left ( \frac{b}{2\mu}\!+\!\frac{a}{4\mu^2}\right )\!(\cosi(1-\workduringckpt)\!-\!\ios)\ccc^2 \; .\\
\end{align*}

Let $\toptE$ be the only positive root of this quadratic polynomial in $T$:  $\toptE$ is 
the value that minimizes \efin.
In the following, we let \algoE be the checkpointing strategy that checkpoints with period \toptE.


As a side note, let us emphasize the differences with the approach of 
Meneses, Sarood and Kal\'e~\cite{Kale2012} when restricting to the case $\workduringckpt=0$ 
(because they only consider the blocking variant). For each failure, they
consider that:
\begin{itemize}
	\item energy lost due to re-execution is $\frac{\period - 2\ccc }{2}\pCal$, while we have
\[ \left( \frac{\period - \ccc}{\period} \left (\frac{\period - \ccc}{2} \right ) + \frac{\ccc}{\period}  \left ( \period - \ccc \right ) \right) \pCal = \frac{\period^2 - \ccc^2}{2\period}  \pCal \]  
	\item energy lost due to I/O is $\ccc \pIO$, while we have $\frac{\ccc^2}{2\period}\pIO$.
\end{itemize}
Theses differences come from our more detailed analysis of the impact of the failure location, which can strike
either during the computation phase, or during the checkpointing phase, of the whole period.

\smallskip
\section{Experiments}
\label{sec.experiments}

In this section, we instantiate the previous model with scenarios taken from current projections
for Exascale platforms~\cite{IESP-Exascale,DARPA,Shalf2011,Ferreira2011}. 
We choose realistic values for all model parameters:
this includes all types of power consumption (\pIdle, \pCal, \pIO and \pDown), all checkpoint parameters
(\ccc, \rrr, \ddd and $\workduringckpt$), and the platform MTBF $\mu$.
We start with a word of caution: our choices for these parameters may be somewhat arbitrary, and 
do not cover the whole range of scenarios that can be investigated. 
However, a key feature of our model is its robustness: as long as
$\mu$ is reasonably large in front of checkpoint times, the model is able to accurately predict the 
best period for execution time and for energy consumption.

The power consumption of an Exascale machine is capped to $20$ Mega-watts. With $10^{6}$ nodes,
this represents a nominal power of $20$ milli-watts per node. Let us express all power values in milli-watts.
A reasonable scenario is to assume that half this power is used for operating the platform,
hence to let $\pIdle = 10$. The overhead due to computing would represent the other half, hence
$\pCal = 10$. As for communications and I/Os, which are expected to cost an order of magnitude
more than computing~\cite{Shalf2011}, we take an overhead of $100$, hence $\pIO =100$.
A key parameter for the experimental study is the ratio 

\begin{equation}
\label{eq.ratio}
\ratio = \frac{\pStatic +\pIO}{\pStatic+\pCal} = \frac{1 + \ios}{1 + \cosi} .
\end{equation}

\begin{figure}[t]
    \centering
    \includegraphics{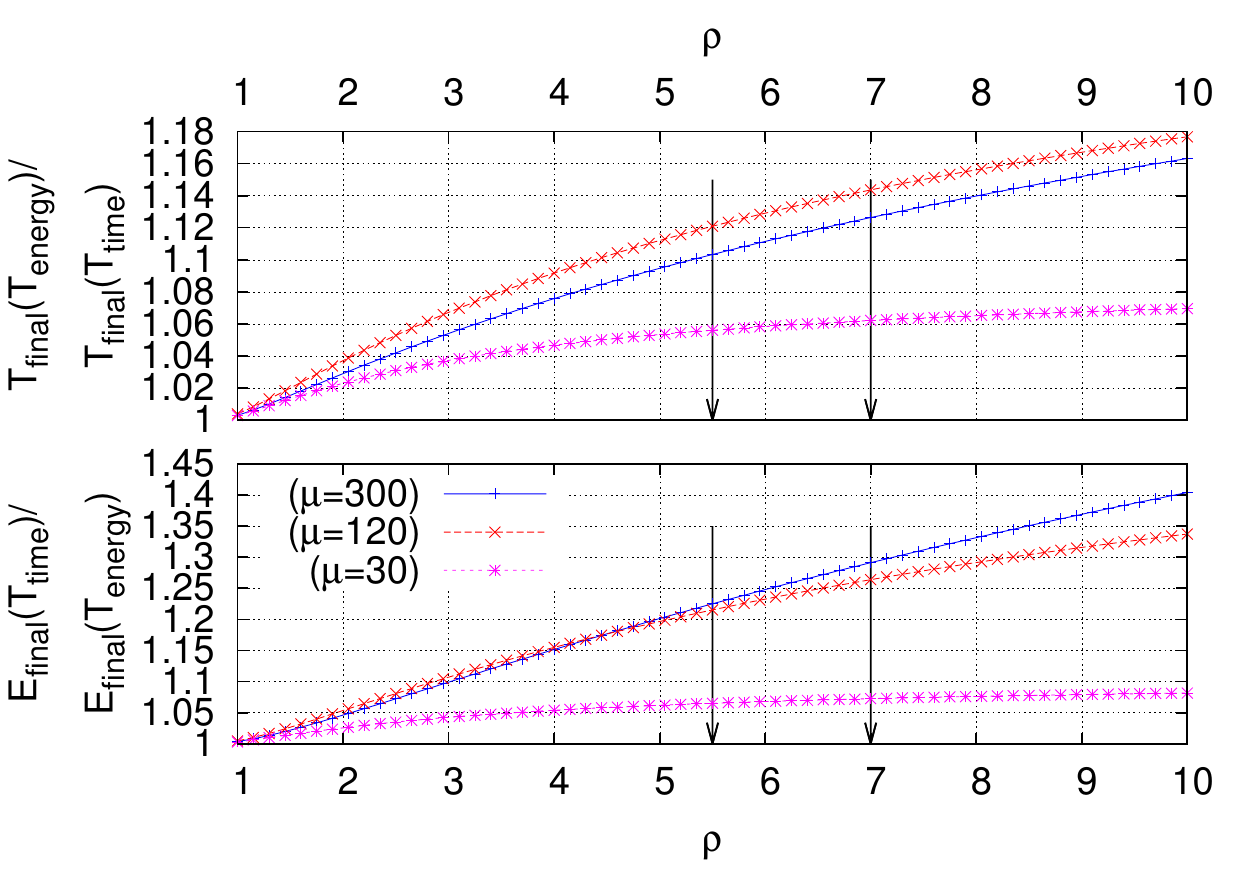}
    \caption{Time and energy ratios as a function of $\ratio$,
      with $C=R=10$ min, $D=1$ min, $\dos=0$, $\workduringckpt=1/2$,
      and various values for $\mu$.}
    \label{fig:energytime}
\end{figure}

With our values,  we get $\ratio=5.5$. 
Note that if we used $\pIdle = 5$ and kept the same overheads
$10$ and $100$ for computing and I/O respectively, we would get $\pCal = 10$,
$\pIO = 100$, and $\ratio =  7$. These two representative 
values of~$\ratio$ ($\ratio = 5.5$ and $\ratio=7$) are emphasized 
by vertical arrows in the plots below on Figure~\ref{fig:energytime}.
As for $\pDown$, the power during downtime, we use $\pDown = 0$, meaning that during downtime we only
account for the static power \pStatic of the processors that are idle.

The Jaguar platform, with $N=45,208$ processors, is
reported to have experienced about one fault per day~\cite{6264677},
which leads to an individual (processor) MTBF $\mu_{\text{ind}}$ equal
to $\frac{45,208}{365}\approx 125$ years. Therefore, we set the
individual (processor) MTBF to $\mu_{\text{ind}} = 125$ years. Letting
the total number of processors $N$ vary from $N=219,150$ to
$N=2,191,500$ (future exascale platforms), the platform MTBF $\mu$ varies 
from $\mu=300$ min ($5$ hours) down to $\mu=30$ min.
The experiments use resilience parameters that are representative of current and forthcoming
large-scale platforms~\cite{Ferreira2011,j116}.  We take $\ccc=\rrr=10$ min,  
$\ddd=1$ min, and $\workduringckpt=1/2$.

\begin{figure}
    \centering
  \begin{subfigure}[b]{0.7\textwidth}
    \centering
    \includegraphics[width=\textwidth]{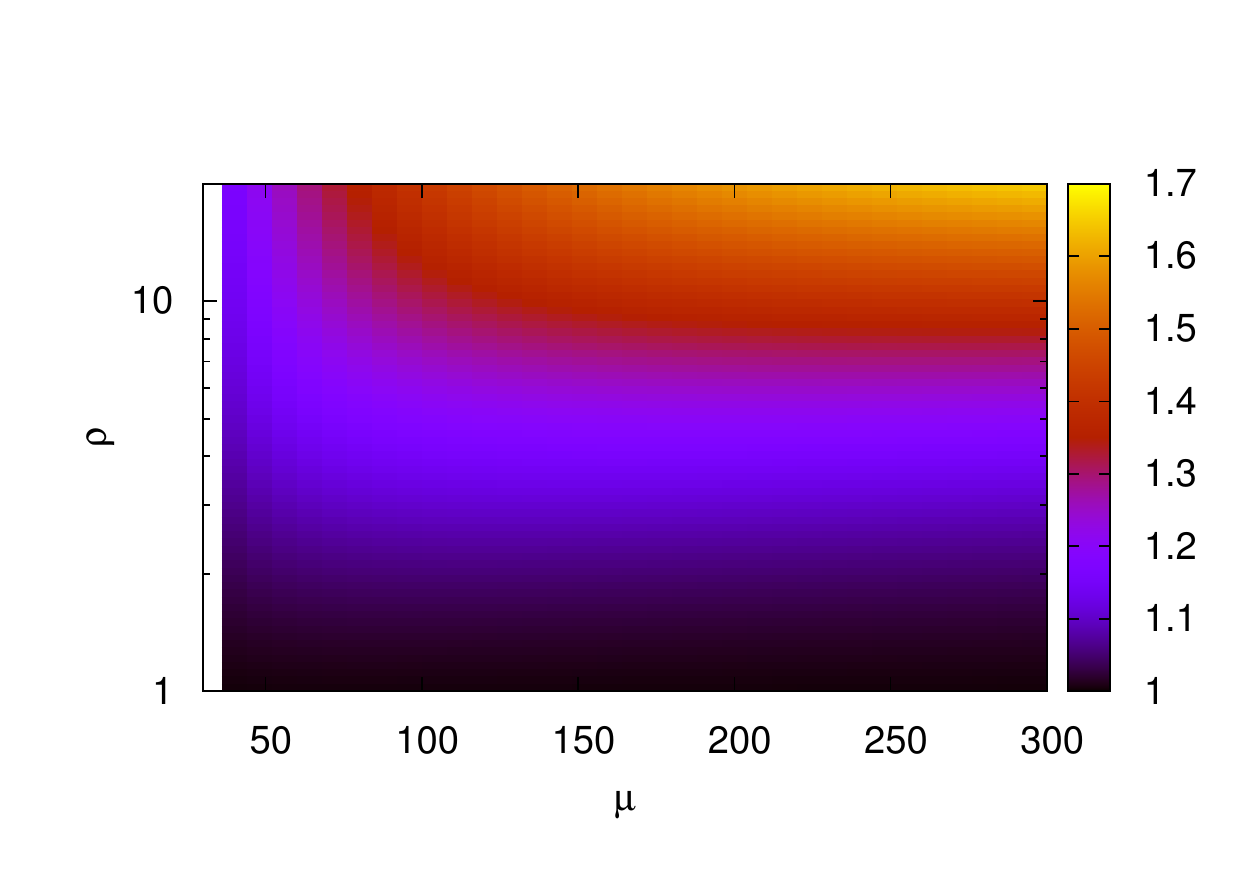}
    \caption{Energy ratio of \algoT over \algoE}
    \label{fig:3d:energy}
  \end{subfigure}\\
  \begin{subfigure}[b]{0.7\textwidth}
    \centering
    \includegraphics[width=\textwidth]{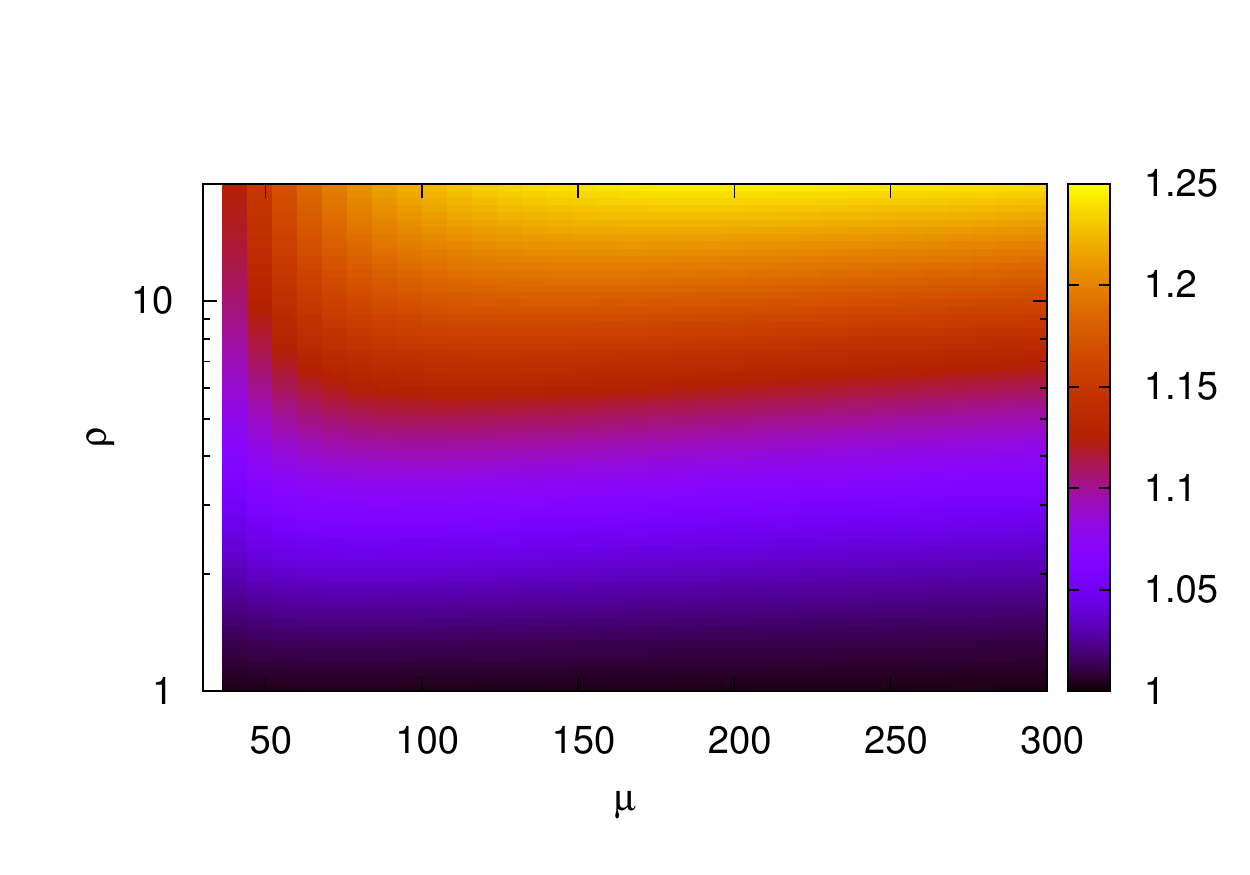}
    \caption{Execution time ratio of \algoE over \algoT} 
    \label{fig:3d:time}
  \end{subfigure}
  \caption{Ratios of the different strategies with $C=R=10$ min, $D=1$ min, $\dos=0$, $\workduringckpt=1/2$
  as a function of $\mu$ and $\ratio$.   \label{fig:3d}}
\end{figure}

On Figures~\ref{fig:energytime} and~\ref{fig:3d}, we evaluate the impact of the 
ratio~$\ratio$ (see Equation~\eqref{eq.ratio}) on the gain 
in energy and loss in time of \algoE with respect to \algoT. The general trend is that using 
\algoE can lead to significant gains in energy at the price of a small increase in execution time.

\begin{figure}
    \centering
  \begin{subfigure}[b]{0.7\textwidth}
    \centering
    \includegraphics[width=\textwidth]{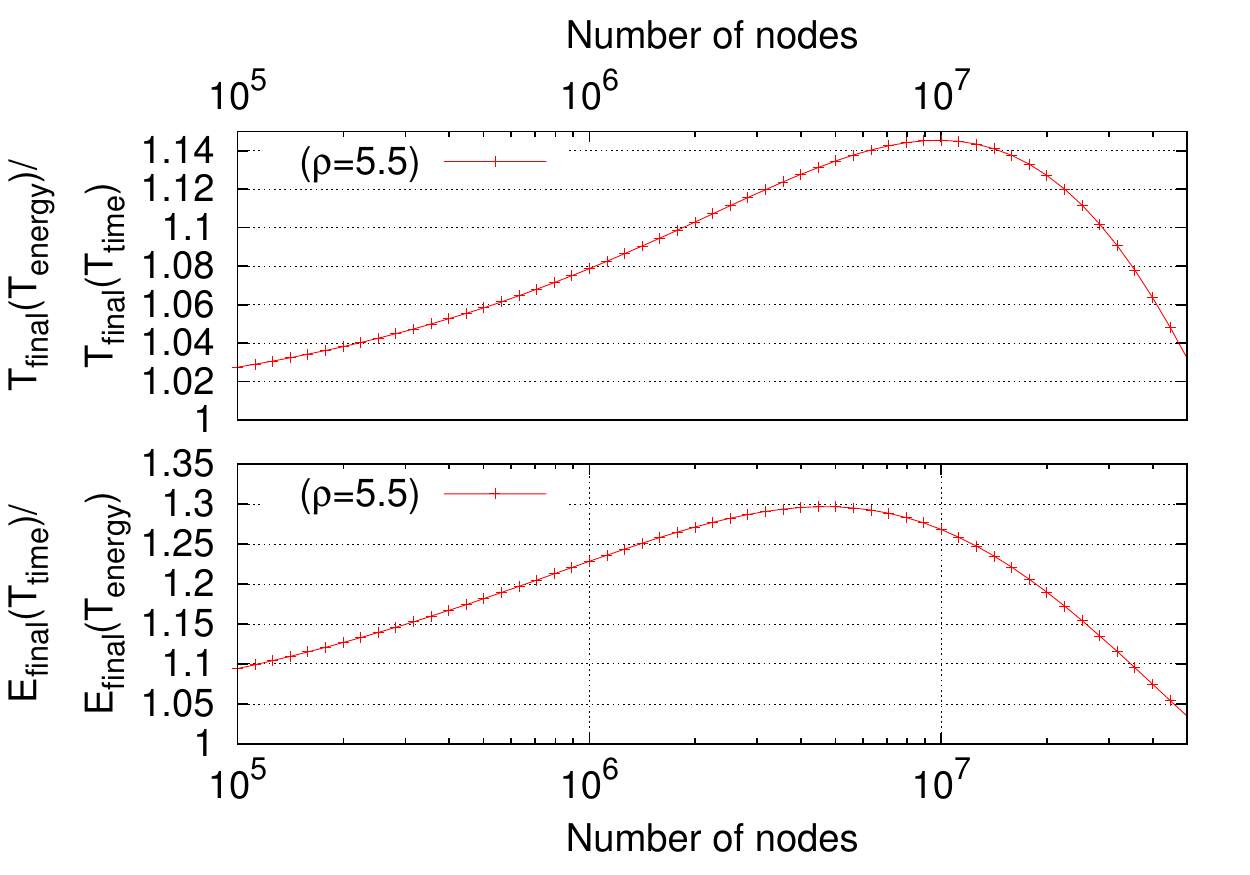}
    \caption{Time and energy ratios, as a function of the
      number of nodes, when $\ratio = 5.5$}
    \label{fig:weak55}
  \end{subfigure}\\
  \begin{subfigure}[b]{0.7\textwidth}
    \centering
    \includegraphics[width=\textwidth]{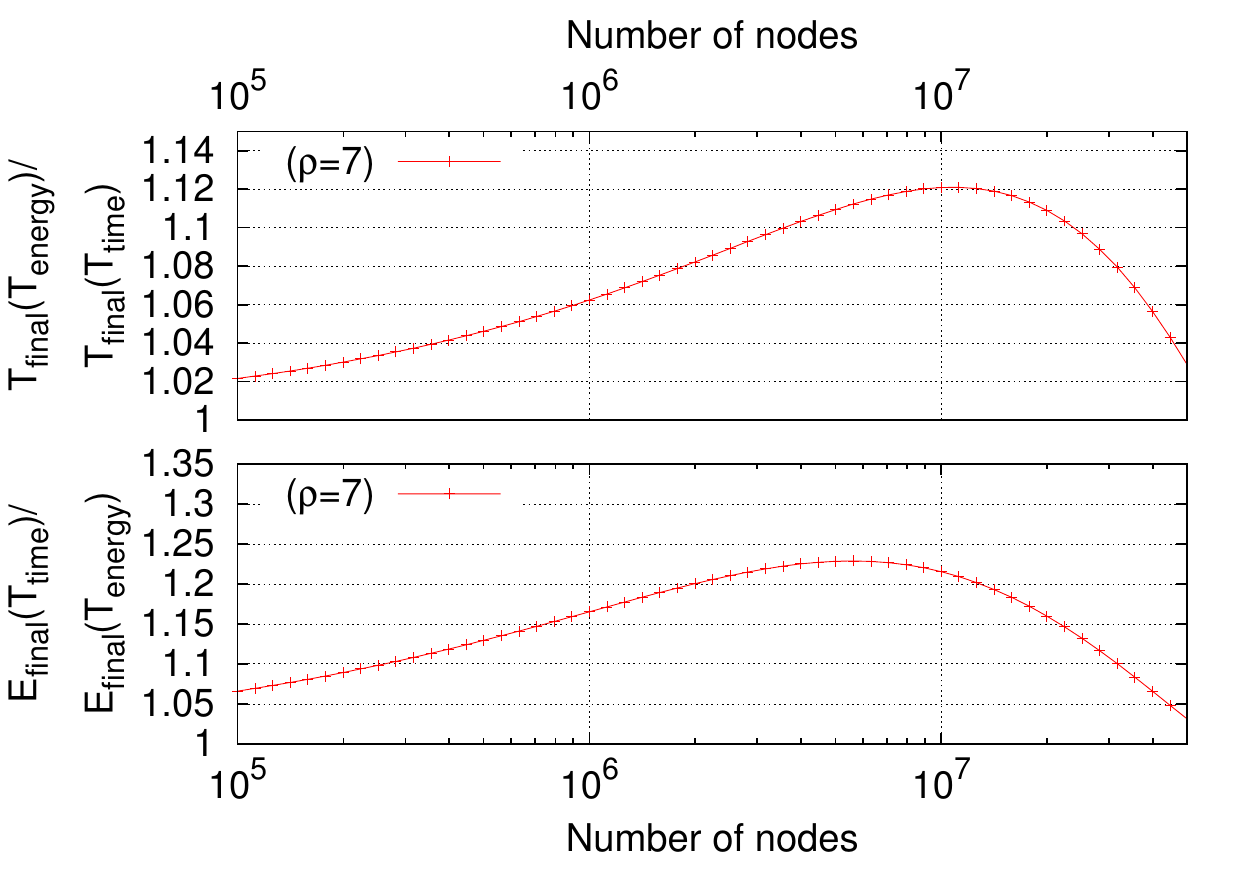}
    \caption{Time and energy ratios, as a function of the
      number of nodes, when $\ratio = 7$}
    \label{fig:weak7}
  \end{subfigure}  
  \caption{Ratios of total energy and time for the two period strategies, as a function of the
      number of nodes, with $\mu = 120$ min 
      for $10^6$ nodes, $C=R=1$ min, $D=0.1$ min, $\dos=0$, $\workduringckpt=1/2$.\label{fig3}}
\end{figure}

We then study in Figure~\ref{fig3} the scalability
of the approach on forthcoming platforms. 
We set the duration of the complete checkpoint and rollback
(\ccc and \rrr, respectively) to $1$ minute, independently of the number of processors,
and we let the downtime \ddd equal to $0.1$ minutes. 
It is reasonable to consider that checkpoint
storage time will not increase with the number of nodes in the future, but on the
contrary will remain constant. Indeed, system designers
are studying a couple of alternative approaches. One consists in
featuring each computing node with local storage capability, ensuring
through the hardware that this storage will remain available during a
failure of the node. Another approach consists in using the memory of
the other processors to store the checkpoint, pairing nodes as
``buddies'', thus allowing to take advantage of the high bandwidth
capability of the high speed network to design a scalable checkpoint
storage mechanism~\cite{ZhengShiKale2004,NiClusterFT12,buddy,OnePetabSecondCheckpoint}.

The MTBF for $10^{6}$ nodes is set to $2$ hours, and this value scales linearly with
the number of components. 
Given these parameters, Figures~\ref{fig:weak55} and~\ref{fig:weak7} 
shows (i) the execution time ratio of \algoE over \algoT, 
and (ii) the energy consumption ratio of \algoT over \algoE, both as a function of the 
number of nodes.
Figures~\ref{fig:weak55} and~\ref{fig:weak7} confirm the important gain in energy that can be achieved,
namely up to $30\%$ for a time overhead of only $12\%$.
When the number of nodes gets very high (up to $10^{8}$), then we observe that both energy and 
time ratios converge to $1$. Indeed, when  \ccc becomes of the order of magnitude of the MTBF,
then both periods \toptT and \toptE become close to \ccc to account for the higher failure rate.

\section{Conclusion}
\label{sec.conclusion}

In this short paper, we have provided a detailed analysis to compute the 
optimal checkpointing period, when the checkpointing activity can be partially
overlapped with computations. We have considered two distinct objectives:
either the goal is to minimize the total execution time, or it is to minimize
the total energy consumption. Because of the different power consumption overheads
due to computations and I/Os, we obtain different optimal periods. 

We have instantiated the formulas with values derived from current and future
Exascale platforms, and we have studied the impact of the power overhead
due to I/O activity on the gains in time and energy. With current values, we can save
more than $20\%$ of energy with an MTBF of 300 min, at the price of an increase of 
$10\%$ in the execution time. The maximum gains are expected for a platform
with between $10^6$ and $10^7$ processors (up to $30\%$ energy savings). 

Our analytical model is quite flexible and can easily be instantiated to investigate scenarios
that involve a variety of resilience and power consumption parameters.

\section*{Acknowledgements} 

This work was supported in part by the ANR {\em RESCUE} project. 
A.~Benoit and Y.~Robert are with the Institut Universitaire de France.

\clearpage
\bibliographystyle{IEEEtran}
\bibliography{biblio}

\begin{thebibliography}{10}
\providecommand{\url}[1]{#1}
\csname url@samestyle\endcsname
\providecommand{\newblock}{\relax}
\providecommand{\bibinfo}[2]{#2}
\providecommand{\BIBentrySTDinterwordspacing}{\spaceskip=0pt\relax}
\providecommand{\BIBentryALTinterwordstretchfactor}{4}
\providecommand{\BIBentryALTinterwordspacing}{\spaceskip=\fontdimen2\font plus
\BIBentryALTinterwordstretchfactor\fontdimen3\font minus
  \fontdimen4\font\relax}
\providecommand{\BIBforeignlanguage}[2]{{%
\expandafter\ifx\csname l@#1\endcsname\relax
\typeout{** WARNING: IEEEtran.bst: No hyphenation pattern has been}%
\typeout{** loaded for the language `#1'. Using the pattern for}%
\typeout{** the default language instead.}%
\else
\language=\csname l@#1\endcsname
\fi
#2}}
\providecommand{\BIBdecl}{\relax}
\BIBdecl

\bibitem{IESP-Exascale}
J.~Dongarra, P.~Beckman, P.~Aerts, F.~Cappello, T.~Lippert, S.~Matsuoka,
  P.~Messina, T.~Moore, R.~Stevens, A.~Trefethen, and M.~Valero, ``The
  international exascale software project: a call to cooperative action by the
  global high-performance community,'' \emph{Int. Journal of High Performance
  Computing Applications}, vol.~23, no.~4, pp. 309--322, 2009.

\bibitem{DARPA}
V.~Sarkar \emph{et~al.}, ``Exascale software study: Software challenges in
  extreme scale systems,'' 2009, white paper available at:
  \url{http://users.ece.gatech.edu/mrichard/ExascaleComputingStudyReports/ECSS%20report%20101909.pdf}.

\bibitem{young74}
J.~W. Young, ``{A first order approximation to the optimum checkpoint
  interval},'' \emph{Comm. of the ACM}, vol.~17, no.~9, pp. 530--531, 1974.

\bibitem{daly04}
J.~T. Daly, ``A higher order estimate of the optimum checkpoint interval for
  restart dumps,'' \emph{FGCS}, vol.~22, no.~3, pp. 303--312, 2004.

\bibitem{Shalf2011}
J.~Shalf, S.~Dosanjh, and J.~Morrison, ``Exascale computing technology
  challenges,'' in \emph{{VECPAR'10}, the 9th Int. Conf. High Performance
  Computing for Computational Science}, ser. LNCS 6449.\hskip 1em plus 0.5em
  minus 0.4em\relax Springer-Verlag, 2011, pp. 1--25.

\bibitem{Kale2012}
E.~Meneses, O.~Sarood, and L.~V. Kal\'e, ``{Assessing Energy Efficiency of
  Fault Tolerance Protocols for HPC Systems},'' in \emph{Proceedings of the
  2012 IEEE 24th International Symposium on Computer Architecture and High
  Performance Computing (SBAC-PAD 2012)}, New York, USA, October 2012.

\bibitem{CL85}
K.~M. Chandy and L.~Lamport, ``Distributed snapshots: Determining global states
  of distributed systems,'' in \emph{Transactions on Computer Systems}, vol.
  3(1).\hskip 1em plus 0.5em minus 0.4em\relax ACM, February 1985, pp. 63--75.

\bibitem{ccpe-2012-ckpt}
G.~Bosilca, A.~Bouteiller, E.~Brunet, F.~Cappello, J.~Dongarra, A.~Guermouche,
  T.~Hérault, Y.~Robert, F.~Vivien, and D.~Zaidouni, ``Unified model for
  assessing checkpointing protocols at extreme-scale,'' \emph{Concurrency and
  Computation: Practice and Experience}, October 2013, to be published. Also
  available as INRIA research report 7950 at \url{graal.ens-lyon.fr/~yrobert}.

\bibitem{Ferreira2011}
K.~Ferreira, J.~Stearley, J.~H.~I. Laros, R.~Oldfield, K.~Pedretti,
  R.~Brightwell, R.~Riesen, P.~G. Bridges, and D.~Arnold, ``{Evaluating the
  Viability of Process Replication Reliability for Exascale Systems},'' in
  \emph{Proc. of the ACM/IEEE SC Conf.}, 2011.

\bibitem{6264677}
G.~Zheng, X.~Ni, and L.~V. Kal\'e, ``A scalable double in-memory checkpoint and
  restart scheme towards exascale,'' in \emph{Dependable Systems and Networks
  Workshops (DSN-W)}, 2012.

\bibitem{j116}
F.~Cappello, H.~Casanova, and Y.~Robert, ``Preventive migration vs. preventive
  checkpointing for extreme scale supercomputers,'' \emph{Parallel Processing
  Letters}, vol.~21, no.~2, pp. 111--132, 2011.

\bibitem{ZhengShiKale2004}
G.~Zheng, L.~Shi, and L.~V. Kal{\'e}, ``{FTC-Charm++: an in-memory
  checkpoint-based fault tolerant runtime for Charm++ and MPI},'' in
  \emph{Proc. 2004 IEEE Int. Conf. Cluster Computing}.\hskip 1em plus 0.5em
  minus 0.4em\relax IEEE Computer Society, 2004.

\bibitem{NiClusterFT12}
X.~Ni, E.~Meneses, and L.~V. Kal{\'e}, ``Hiding checkpoint overhead in {HPC}
  applications with a semi-blocking algorithm,'' in \emph{Proc. 2012 IEEE Int.
  Conf. Cluster Computing}.\hskip 1em plus 0.5em minus 0.4em\relax IEEE
  Computer Society, 2012.

\bibitem{buddy}
J.~Dongarra, T.~Hérault, and Y.~Robert, ``Revisiting the double checkpointing
  algorithm,'' in \emph{15th Workshop on Advances in Parallel and Distributed
  Computational Models {APDCM 2013}}.\hskip 1em plus 0.5em minus 0.4em\relax
  IEEE Computer Society Press, 2013.

\bibitem{OnePetabSecondCheckpoint}
R.~Rajachandrasekar, A.~Moody, K.~Mohror, and D.~K.~D. Panda, ``{A 1 PB/s file
  system to checkpoint three million MPI tasks},'' in \emph{Proceedings of the
  22nd international symposium on High-performance parallel and distributed
  computing}, ser. HPDC '13.\hskip 1em plus 0.5em minus 0.4em\relax New York,
  NY, USA: ACM, 2013, pp. 143--154.

\end{thebibliography}
\end{document}